\documentclass[onecolumn,preprint,floatfix,amsmath,amssymb]{revtex4}

\usepackage{graphicx}
\usepackage{dcolumn}
\usepackage{bm}
\usepackage{subfigure}
\begin{document}
\newcommand{\kB}{k_{\mathrm{B}}}
\newcommand{\me}{\mathrm{e}}
\newcommand{\K}{\mathrm{K}}
\newcommand{\diff}{\mathrm{d}}
\newcommand{\br}{\textbf{r}}
\newcommand{\rur}{\mathrm{Ru}^{2+}}
\newcommand{\ruo}{\mathrm{Ru}^{3+}}
\newcommand{\cl}{\mathrm{Cl}^-}


\title{Water at an electrochemical interface - a simulation study}

\author{Adam P. Willard$^2$}
\author {Stewart K. Reed$^1$}
\author{Paul A. Madden$^1$}
\author{David Chandler$^2$}
\affiliation{$^1$School of Chemistry, University of Edinburgh, Edinburgh EH9 3JJ, UK}
\affiliation{$^2$Department of Chemistry, University of California, Berkeley,California 94720}



\date{\today}

\begin{abstract}
The results of molecular dynamics simulations of the properties of water in an
aqueous ionic solution close to an interface with a model metallic electrode are
described. In the simulations the electrode behaves as an ideally polarizable
hydrophilic metal, supporting image charge interactions with charged species, and
it is maintained at a constant electrical potential with respect to the solution so
that the model is a textbook representation of an electrochemical interface through
which no current is passing. We show how water is strongly attracted to and ordered
at the electrode surface.  This ordering is different to the structure that might
be imagined from continuum models of electrode interfaces. Further, this ordering
significantly affects the probability of ions reaching the surface.  We describe
the concomitant motion and configurations of the water and ions as functions of the
electrode potential, and we analyze the length scales over which ionic atmospheres
fluctuate.  The statistics of these fluctuations depend upon surface structure and
ionic strength. The fluctuations are large, sufficiently so that the mean ionic
atmosphere is a poor descriptor of the aqueous environment near a metal surface.
The importance of this finding for a description of electrochemical reactions is
examined by calculating, directly from the simulation, Marcus free energy profiles
for transfer of charge between the electrode and a redox species in the solution
and comparing the results with the predictions of continuum theories. Significant
departures from the electrochemical textbook descriptions of the phenomenon are
found and their physical origins are characterized from the atomistic perspective
of the simulations.
\end{abstract}

\maketitle
\section{\label{sec:introduction}Introduction}
The altered solvation, dielectric and dynamical properties of water molecules close
to electrode surfaces have an important influence on electrochemical reactions.
There have been numerous simulation studies of aqueous solutions close to charged
solid surfaces which have cast light on the ordering of water molecules by the
solid surface and begun to make the connection between the layers with altered
dielectric characteristics invoked in continuum models of the electrode capacitance
and the ordered molecular films which are known from surface science. References
\cite{neurock} contain an excellent summary of work to date, and
\cite{neurock1,spohr} a more longstanding review. To complete the link between
molecular behaviour and electrochemical observations we need a realistic
representation of the electrochemical interface and a direct way of calculating the
electrochemical observable, namely the dependence of the rate of the
electrochemical electron transfer on the potential applied to the electrode
\cite{bockris}.

\begin{figure}
\includegraphics[width=4.5in]{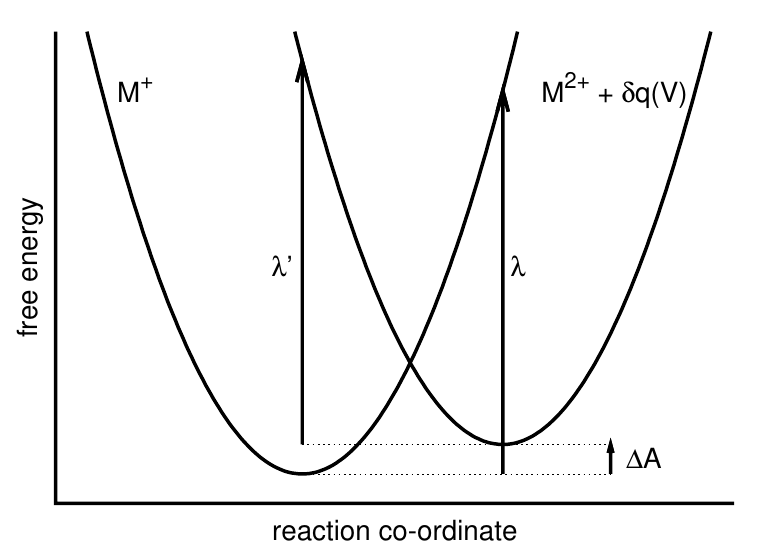}
\caption{Schematic illustration of the Marcus construction for understanding the
electrochemical oxidation of some species R= M$^+$ to O=M$^{2+}$ with  transfer of
the charge ($\delta q$) to an electrode maintained at a potential $V$. The
quantities $\Delta A$, $\lambda$ and $\lambda^\prime$ are the reaction free-energy
and reorganization energies for oxidation and reduction, respectively}
  \label{fig:marcustheory}
\end{figure}

What is required to achieve  the second of these objectives is suggested by the
Marcus theory of electron transfer \cite{marcus1,marcus2}. In a Marcus description
of the oxidation of some solution species R to an oxidised species O by transfer of
an electron to an electrode maintained at some potential $V$ with respect to the
solution we construct curves, as illustrated schematically in figure
\ref{fig:marcustheory}, which describe how the free energies of O and R depend upon
some reaction coordinate, which is envisaged as reflecting the influence of
fluctuating solvent degrees of freedom. The Marcus expression for the rate of
electron transfer can be calculated from the probability that the system will
access the configuration where the two curves cross. Note that the free energy
curve for O includes the potential energy of the electron on the electrode ($eV$),
so that the position and height of the crossing point depend on the electrode
potential. The two curves are coupled by a term ($\gamma$) which reflects the
tunneling of the electron between the redox centre and the electrode, and this is
expected to depend exponentially on its distance from the electrode surface. We
should, therefore, be thinking about the dependence of the Marcus curves on the
proximity to the electrode surface, to which two factors contribute. Firstly, the
\emph{difference} between the \emph{direct} interactions of the O and R species
with the charged surface itself will produce a differential shift on them, and
therefore affect the crossing point. Secondly, if the redox species is close to the
electrode, the competing interactions of the water molecules in its coordination
shell with the surface and the solute itself may result in a change in the
character of the fluctuations of the reaction coordinate. Both of these factors may
be affected by the potential applied to the electrode. The Marcus curves contain
the information which is required to calculate the electron transfer rate for a
redox ion at a given distance from the electrode surface, but to complete the
calculation of the rate we also need to know the probability of the reactant
reaching this position, and this too will be affected by the potential exerted by
the electrode and its influence on the solvating properties of the water molecules.

\begin{figure}[htbp]
  \centering
 \includegraphics[width=5in]{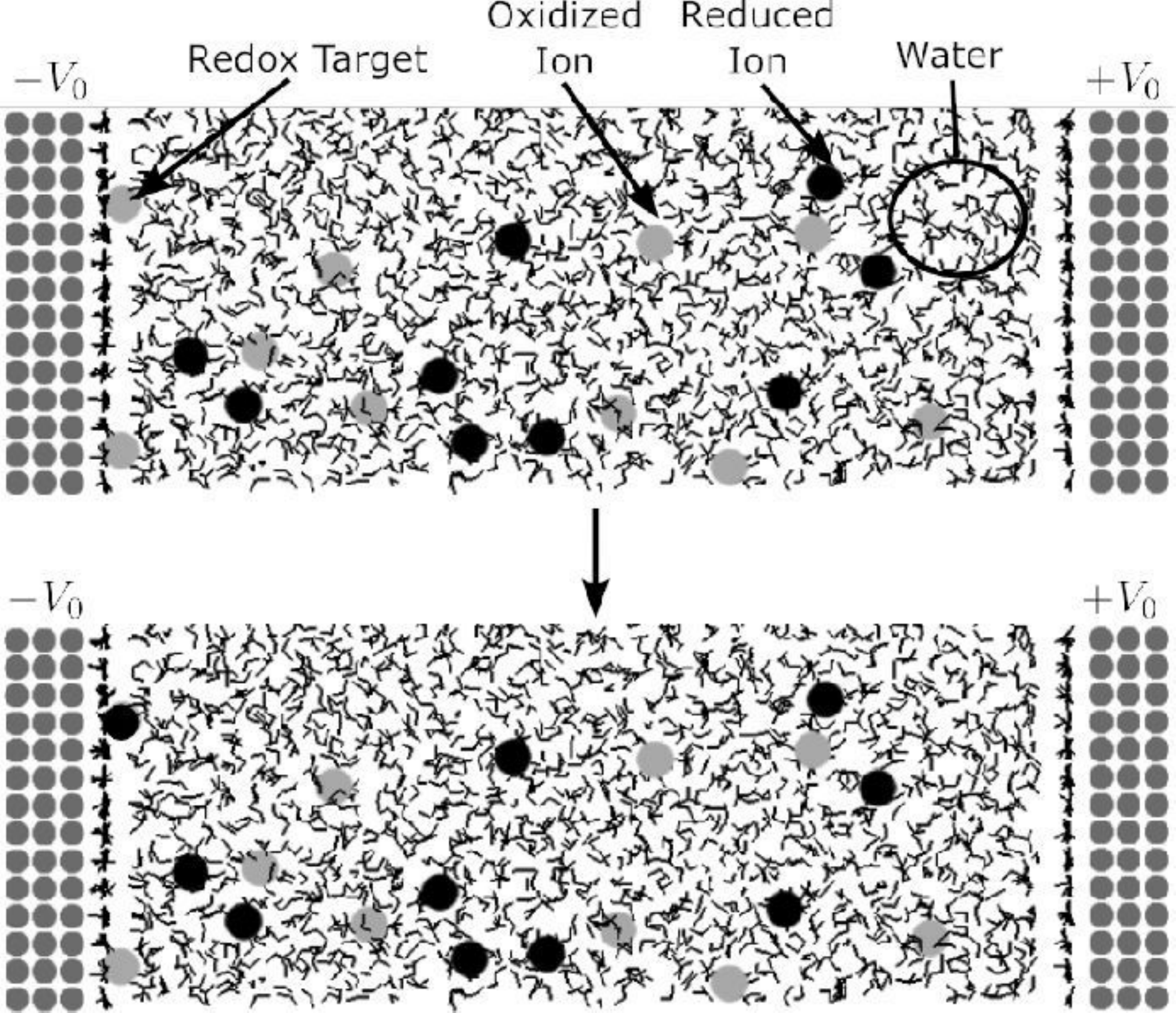}
  \caption{Schematic depiction of electron transfer}
  \label{fig:cell}
\end{figure}
Blumberger, Sprik and co-workers \cite{sprik1,sprik2} have demonstrated how the
Marcus curves can be calculated for a homogeneous electron transfer reaction within
an {\it ab initio} molecular dynamics scheme. Following Warshel \cite{warshel} they
emphasize the advantages for computation of choosing the ``vertical energy-gap" as
the reaction coordinate. The vertical energy gap for oxidation is calculated for a
single configuration in a simulation by switching the identity (and all associated
interaction parameters) of a redox species initially in its reduced form R to its
oxidised form without allowing any changes in the nuclear coordinates (hence a
``vertical" transition in the Franck-Condon sense) and evaluating the energy
difference between the final and initial states. The Marcus curves for the oxidised
and reduced species may then be estimated from the probability distributions of the
vertical energy gaps obtained by repeatedly sampling through the course of an
molecular dynamics (MD) simulation and assuming this distribution is Gaussian.  The
Gaussian assumption is required by this method because regions of the distribution
pertinent to the charge transfer reaction are not generally accessed in a
straightforward simulation.  In the calculations reported here, the Gaussian
approximation  is tested and shown to be accurate.  More generally, the Gaussian
approximation has been tested and found to be accurate~\cite{chandler1,chandler2}
provided proper account is taken of molecular boundary conditions~\cite{chandler3}

Use of this scheme to study the electrochemical
electron transfer process in a simulation is illustrated in figure \ref{fig:cell}.
An aqueous solution is contained between two crystalline arrays of atoms which
comprise the (metallic) electrodes, these are maintained at a definite electrical
potential.  The solution contains the redox species in its reduced and oxidised
forms and, periodically during the simulation, an ion  (the ``redox target") is
selected and its redox state is switched and the energy difference between final
and initial states is evaluated. By selecting ions at different distances from the
electrodes and by examining how the vertical energy gaps depend on the applied
potential we can build up the necessary information to study how the nature of the
water at the electrode surface affects the electrochemical electron transfer rate
{\it via} the Marcus construction.

In order to make contact with experimental studies the calculation needs to be done
with as realistic representation of the constant-potential electrode and
interfacial water as can be managed. Ideally, it would be done within an {\it ab initio} MD
 scheme, as this would enable the difficult-to-characterise interactions
between the solution and the electrode to be modelled without the introduction of
interaction potentials \cite{neurock,halley,voth}. However, the time and length
scales involved in the relaxation of the solution in the vicinity of the electrodes
(which we will characterise below) are far too long to allow a full self-consistent
description of the screening of the electrode potential within an {\it ab initio}
scheme \cite{halley-acs}. Recently we introduced a way of incorporating some of the
essential physical effects necessary for a realistic description of the interfacial
charge-transfer process into a simulation which uses interaction potentials and
therefore enables simulations of much larger time and length scales than is
possible {\it ab initio} \cite{reed1}. In particular, model \emph{metallic}
electrodes maintained at a \emph{constant electrical potential} may be introduced
into such a simulation, following a technique introduced by Siepmann and Sprik
\cite{siepmann}. Because the electrodes behave as ideally polarizable metals they
support image-charge interactions between charged species and the electrode; these,
as we shall see, have an important influence on the electron transfer process.
Because the electrodes are maintained at a constant potential when the charge of
the redox species is changed to sample the vertical energy gap, that charge is
transferred in full to the electrodes, so that the source of the dependence of the
electron-transfer rate on the  electrode potential is included in the calculation.
The electrode potential and the potential felt by the molecules and ions in the
solution region are calculated self-consistently. Calculations using these methods
have already been performed to examine the Marcus curves in simulations of redox
active molten salts \cite{reed2}.

We begin with a brief description of the methods and interaction potentials used to
simulate pure water and aqueous solutions of LiCl and the Ru$^{2+}$/Ru$^{3+}$
couple confined between model platinum electrodes. We then examine the structure
and dynamical properties of the electrode-adsorbed water and the way they are
affected by the application of a potential to the electrode. In sections
\ref{water} and \ref{ions} we consider the consequences of this adsorbed water for
the approach of ions to the electrode surface and the effect of the adsorbed water
and the ionic atmosphere for the electrical potential in the vicinity of the
electrode. In classical models of electrochemical charge transfer this potential is
invoked to represent the dependence of the energies of the oxidised and reduced
species on the proximity to the electrode. We then present preliminary results for
the Marcus curves for the Ru$^{2+}$/Ru$^{3+}$ system and discuss the physical
factors which determine their dependence on the applied potential and the proximity
to the electrode.

\section{\label{sec:model}Scope of the model.}
The electrochemical interface is affected by many phenomena and a comprehensive
representation of all of them within a single simulation is beyond current
capabilities. Our focus here is on the solution side of the interface, on the
properties of the water molecules at the interface and their influence on the
electrical potential. As such we will present a significantly simplified model of
the electrode itself in which we ignore the motion of the electrode atoms and
therefore neglect effects like the restructuring of the electrode surface under
chemical or electrical influences \cite{kolb}. Furthermore the representation of
the electrode as a metal is a simplified one, designed to capture the correct
macroscopic response to an electrical potential appropriate to a metal rather than
to deal with a correct microscopic description of surface electronic states {\it
etc.}
 \begin{figure}[htbp]
  \centering
 \includegraphics[width=5in]{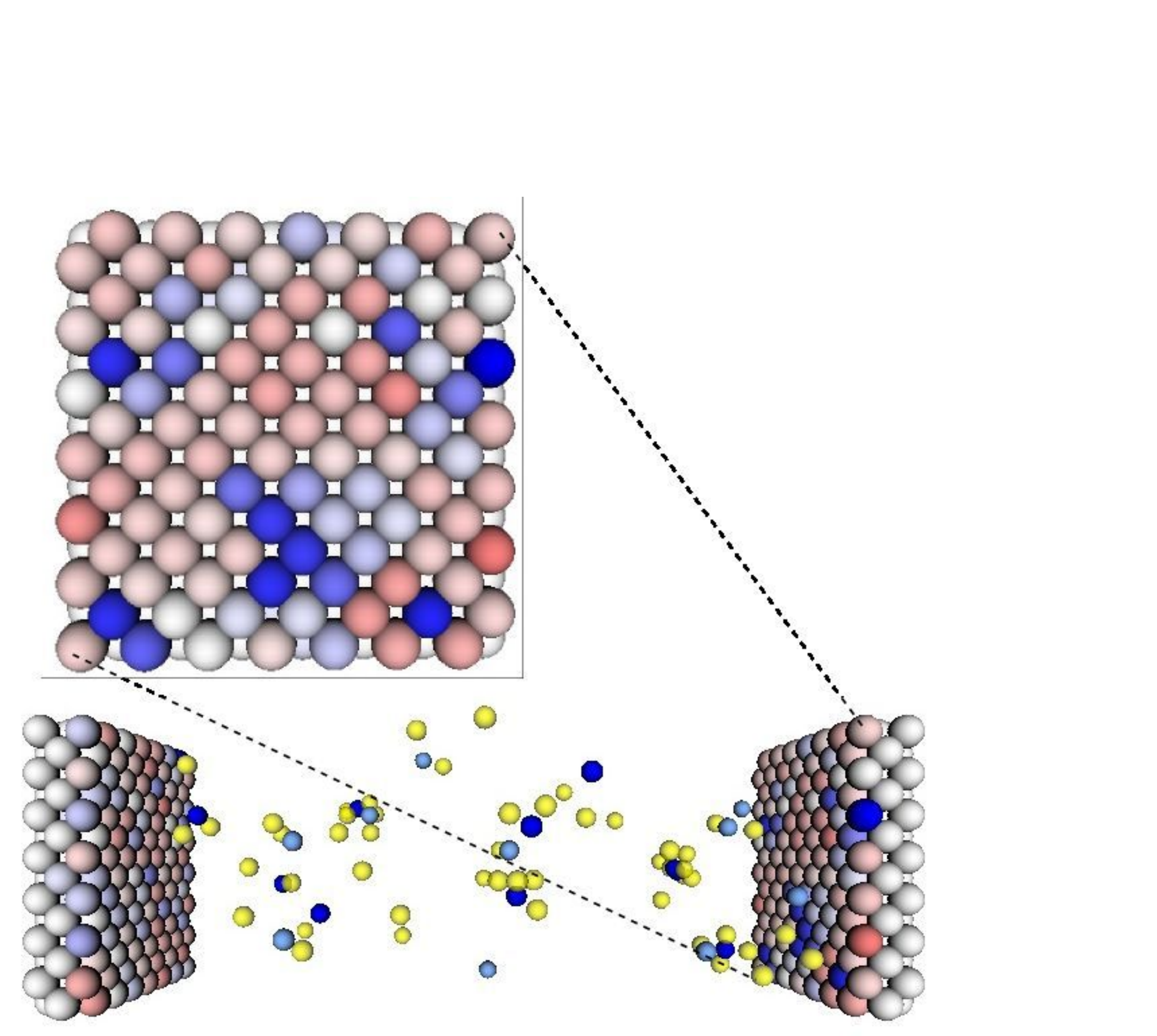}
  \caption{Snapshot of the distribution of charge on the polarizable model electrode.
Electrode atoms are shaded according to their partial charge, with blue shading
corresponding to positive charge and red shading corresponding to negative charge.
In this rendering, the darkest shade of blue corresponds to a partial charge of
$0.1e$, and the darkest shade of red to a partial charge of $-0.1e$.
In the solvent region the yellow spheres represent $\mathrm{Cl^-}$ ions, and the light blue and dark blue spheres represent $\mathrm{Ru^{2+}}$ and $\mathrm{Ru^{3+}}$ respectively.  Water molecules have been omitted for clarity.}
 \label{fig:images}
\end{figure}

As illustrated in figures \ref{fig:cell} and \ref{fig:images}, the electrodes each
consist of three layers of atoms arranged in an fcc lattice with the 100 face
exposed to the solution; the lattice parameter is appropriate to Pt. Following
Siepmann and Sprik \cite{siepmann} each electrode atom $i$ carries a Gaussian
charge distribution of fixed width but variable amplitude ($q_i$). These charges
are coulombically coupled to all other charges in the system. They are treated as
additional dynamical degrees of freedom whose values are adjusted at each timestep
in the molecular dynamic procedure in order to variationally minimise an
appropriate energy functional. The energy functional is chosen \cite{siepmann} so
that at its minimum the electrical potential on every electrode atom is the same
(as approriate to a metal) and equal to some pre-set value $V_0$. The use of a
variational principle allows forces and response behaviour to be calculated
straightforwardly via an application of the Hellmann-Feynman Theorem, as used to
good effect in {\it ab initio} MD simulations. The methodology for
the simulation of the electrodes is described in great detail in ref \cite{reed1},
where it is shown that the electrodes become polarised in the presence of a charge
in the solution region in a way which corresponds to the classical image-charge
response. We illustrate this response in figure \ref{fig:images} where the charge
induced on the electrode atoms by the instantaneous configuration of the charges in
solution is shown by colour-coding the electrode atoms. The bright blue region, for
example, is caused by the presence of an anion within the first  molecular layer of
the solution close to this position. Furthermore, if the charge on one of the ions
in solution is changed, the charge difference is fully transferred to the
electrodes to maintain charge neutrality with a constant electrode potential
\cite{reed2}: it is this feature which enables us to calculate Marcus curves for
electrochemical charge transfer.

To describe the interactions between water molecules we use the SPC/E potential
\cite{SPCE}, which is known to give a dielectric constant for water close to the
experimental value. The interactions of water molecules with a metallic electrode
are complex, and cannot be modelled accurately with a simple two-body potential;
since the behaviour of water at the interface is the central purpose of our study
we were concerned to represent this interaction as carefully as is possible through
the introduction of a potential. Experiment \cite{thiel} and ab-initio studies
\cite{holloway}  have shown that water molecules interact with a crystalline
platinum surface by adsorbing on top sites and orienting their dipole along the
plane of the electrode. In their study of water at an STM tip, Siepmann and Sprik
\cite{siepmann} parameterized a two- and three-body potential to describe the
adsorption of water molecules on a platinum surface; their potentials are
particularly appropriate for our study since they do not include the consequences
of image interactions, which are dealt with through the polarizable electrode model
as in our simulations. We have used these potentials exactly as described in their
paper.

We have not attempted the same level of realism with the interactions between the
ions and the electrode surface. Experimental studies show that anions interact
quite strongly with transition metal surfaces to the extent that complete surface
coverage of ordered layers of Cl$^-$ is observed on positively charged
single-crystal electrodes above about 0.5 V from molar solutions \cite{magnussen}.
Guymon {\it et al} have shown how suitable potentials to describe these strong
interactions could be obtained from {\it ab initio} calculations \cite{guymon}.
However, adsorption of this strength would present a significant problem for our
simulations since it would mean that the solution region would be strongly depleted
in Cl$^-$ ions. To represent the interface under these conditions we would need to
equilibrate our system in the presence of a reservoir of electrolyte; furthermore
equilibrating this system would be \emph{very} slow, as we shall see. We have,
therefore, for the present study introduced only weakly attractive interaction
potentials between the ions and the electrode surface -- we use exponential-6
potentials to represent the short-range interactions with the parameters chosen as
if the atoms of the metallic walls were themselves Cl$^-$ ions. These potentials
are too weakly attractive, compared to the water-electrode interactions, to lead to
the kind of anion adsorption phenomena seen in the experimental studies of
Cl$^-$-containing electrolytes. We note that fluoride ions are not thought to form
adsorbed layers \cite{magnussen} and so the picture of the interface we present may
be more representative of a fluoride than a chloride-containing solution.

The interactions between the other species present in solution were modelled with
pair potentials; this too reflects a compromise in the realism of the calculations
as it is known that polarisation effects have a significant influence on the way
that ions interact and coordinate water \cite{polarizable-ion-water}. The water-ion
interactions were modelled with a Lennard-Jones potential acting between the ion and
the oxygen center of the water molecule.  The parameters used in these interactions
were adapted from Lynden-Bell \cite{bell1}. The ruthenium ion - water interactions
were parameterized with a purely repulsive potential,
\begin{equation}
U_{\mathrm{Ru},\mathrm{O}} = A/\vert \br_\mathrm{Ru} - \br_\mathrm{O} \vert ^9,
\end{equation}
with the parameter ($A = 49977.9 k\mathrm{J mol^{-1} \AA^9}$ for both Ru$^{2+}$ and
Ru$^{3+}$) chosen so that the first peaks of the ion-water radial distribution
agreed with those obtained in an {\it ab initio} MD study \cite{sprik_ru}. The ions
interact with each other with exponential-6 potentials with the Ru-Cl potentials
taken from lanthanides of corresponding ionic size \cite{morgan,hutchinson}.

\section{Pure water results}
\label{water}

\begin{figure}[htbp]
  \centering
   \includegraphics[width=5in]{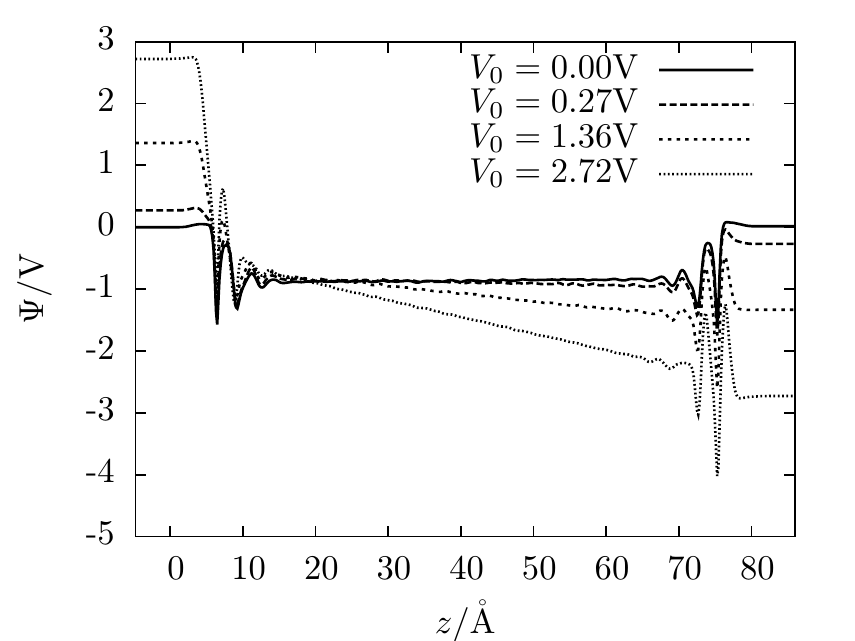}
  \caption{The Poisson potential across the model electrochemical cell for pure water.
    The oscillations near the electrodes result from ordering of the solvent
    in the vicinity of the electrode.}
  \label{fig:poisson1}
\end{figure}
We begin by showing, in figure \ref{fig:poisson1}, results obtained for the
profiles across the cell of the mean electrical (or ``Poisson") potential, $\Psi$,
in pure water. This is obtained by integrating Poisson's equation,
\begin{equation}
    \nabla^{2}\Psi = -\frac{\rho}{\varepsilon_{0}} \;,
    \end{equation}
with the mean charge density, $\rho$, calculated from the simulation for different
values of the applied electrode potential, $V_0$, as the source term
($\varepsilon_{0}$ is the permittivity of free space). The Poisson potential is the
potential used in describing the potential at the electrochemical interface in
classical theories and is therefore an important point of contact between our
calculations and textbook descriptions of the electrochemical interface
\cite{bockris,kornyshev}. The potential is constant on the interior of the
electrodes and equal to the applied potential. It then drops rapidly and oscillates
across an interfacial region about 12 \AA{} wide, for reasons we will discuss in
detail below, before settling down to acquire the constant slope appropriate to the
behaviour of the potential in a bulk dielectric subject to an external potential.
Notice that, other than at $V_0=0$, the potential drops across the two interfaces
are not symmetrical because of the different microscopic arrangements of the water
molecules at positively and negatively charged surfaces.

We can calculate a value for the dielectric constant of water from the behaviour of
the potential across the bulk region. By integrating the mean charge in the region
of the cell to the left of 20 \AA~ we can obtain a value for the charge, Q, on one
plate of a virtual parallel plate capacitor placed at this position; the region to
the right of 61.5 \AA~ has an equal and opposite charge and can be regarded as the
other plate. The potential drop between the two plates is $\Delta \Psi$ and we can
obtain values for the capacitance $C$ from $Q=C\Delta \Psi$ for the different
applied potentials. This calculated capacitance $C$ can be compared with
theoretical expression for capacitance of parallel plates filed with a medium of
dielectric constant $\varepsilon$,
\begin{equation}
C=\varepsilon\varepsilon_0 A/d,
\end{equation}
\begin{figure}[htbp]
  \centering
 \includegraphics[width=6in]{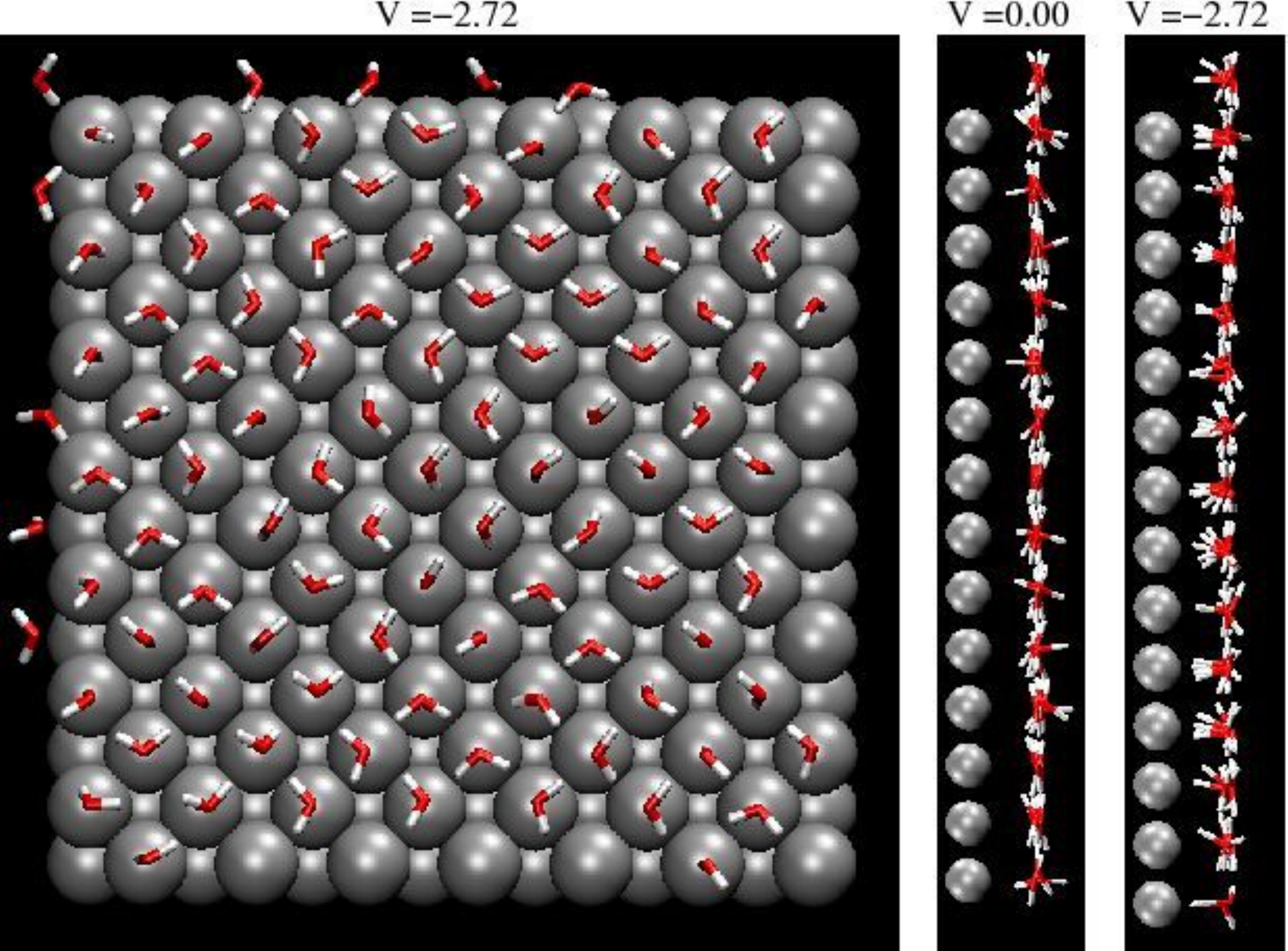}
  \caption{A snapshot of the adsorbed water layer, and their profiles at different
   values of the applied potential $V_0$.
     The orientation of adsorbed water molecules is noticeably
      altered by the change in electrode potential.}
  \label{fig:orient}
\end{figure}
where $A$ is the cross-sectional area of the cell and $d$ the distance between the
plates. The calculation shows that the dielectric constant for SPC/E water depends
on the applied electrode potential.  Specifically, when the electrode potential has
the values $V_0 = 0.27$ V, $V_0 = 1.36$ V, and $V_0 = 2.72$ V, the dielectric
constant for the bulk water is $\epsilon = 75.07$, $\epsilon = 61.50$, and
$\epsilon = 57.30$ respectively.  The low voltage result is in reasonable agreement
with direct simulation studies (68$\pm$5.8 \cite{steinhauser}), which is good
confirmation that the potential and the response of the water molecules to it are
correct (see also reference \cite{reed1}).  The reduction in the apparent
dielectric constant at higher voltages is consistent with a saturation effect
\cite{berkowitz}, note that the potential difference of $\Delta \Psi\simeq$ 1.1 V,
obtained with $V_0$=2.72 V, across our virtual capacitor of width 41.5 \AA~ is
equivalent to an electric field of $\simeq 2.6 \times 10^8$ Vm$^{-1}$.

The rapid oscillation of the potential close to the interfaces is due to the strong
adsorption of a layer of water molecules at the electrode surfaces. This is
illustrated in figure \ref{fig:orient}. The oxygen atoms of the water molecules
form a commensurate layer on the top-sites of the underlying 100 fcc surface. With
zero applied potential the water molecules lie with at least one O-H bond in the
plane of the interface with the H-atom pointing towards a neighbouring oxygen and
form ordered domains which reorient on a long timescale. Although the predominant
orientation is in-plane, at zero applied potential there is a net orientation of
the negative ends of the molecular dipoles towards the surface and this induces a
small positive charge on the electrode atoms \cite{berkowitz}.

When the potential is applied to the cell, the water molecules in the first layer
partially reorient in the interfacial field. This is illustrated in the right-most
panel of figure \ref{fig:orient} where one of the OH bonds of the water molecules
at the negatively charged electrode may point towards the electrode surface (note
that this potential is very large for aqueous electrochemistry). The consequences
of this for the Poisson potential can be see by reference to the right-hand
interface in  figure \ref{fig:poisson1}. Whereas at $V_0=0$ the potential initially
drops on moving into the electrolyte, consistent with the negatively charged oxide
ions being closest to the surface, at the applied potential of $V_0=-2.72$ V, the
potential now rises showing an excess of positive charge lying close to the
surface.

\begin{figure}
   \includegraphics[width=7 cm]{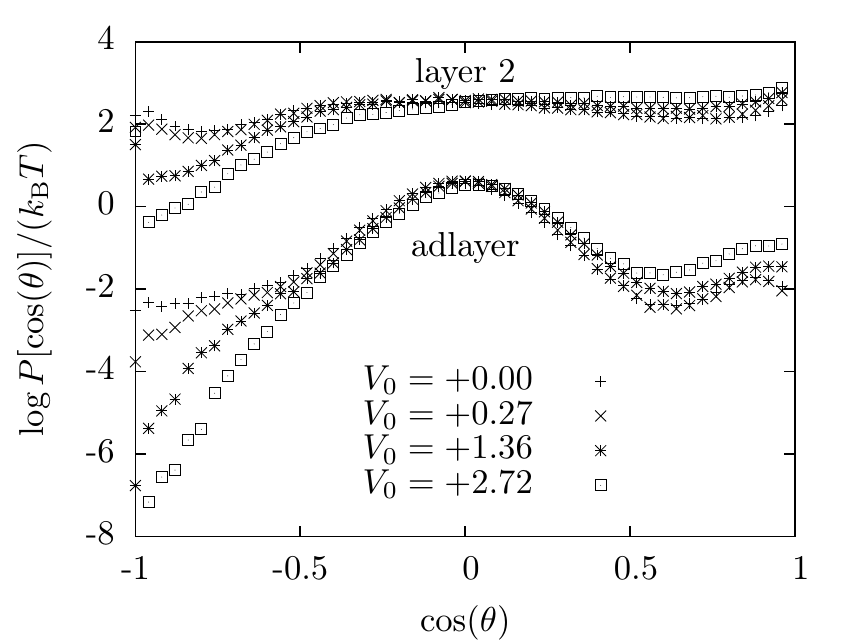}\includegraphics[width=7 cm]{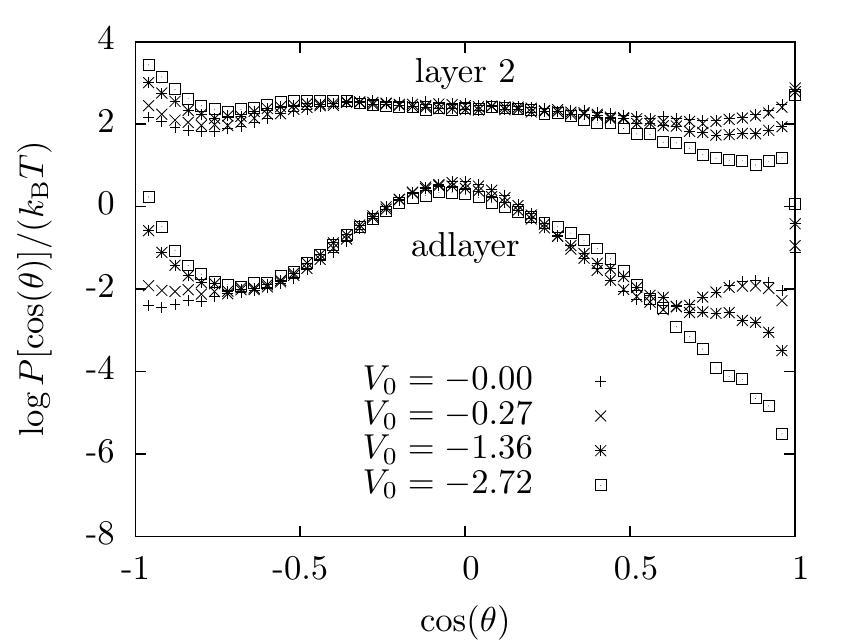}
\caption{The distribution of $\cos(\theta)$ where $\theta$ is the angle between a
water's O-H bond vector and the vector normal to the electrode surface (pointing
into the bulk).  The left-hand panel is the distribution taken over the first two
solvent layers adjacent to the positive electrode, and the right-hand one the
distribution taken over the first two solvent layers adjacent to the negative
electrode.}
  \label{fig:OHorient}
\end{figure}

We can track these changes by showing the probability distributions of the
orientation of O-H bonds in the first and second layers of water molecules as the
applied potential is changed. We compute the probability distribution function
$P[\cos (\theta)]$, where $\theta$ is the angle between an OH bond vector (the
vector extending from the oxygen centre of a water molecule to the centre of one of
the associated hydrogen atoms) and the the outward normal vector to the electrode
surface.  Fig \ref{fig:OHorient} shows the distribution $P[\cos(\theta)]$ for
different values of applied potential for the two electrodes. The left-hand panel of
figure \ref{fig:OHorient} shows the results at the positively charged electrode
and the right-hand panel at the negatively charged one. At all values of the potential, the
probability distribution is peaked around $\cos(\theta) = 0$, corresponding to
configurations for which the OH vector of a water molecule is aligned with  the
plane of the electrode surface. At zero applied potential the distribution
$P[\cos(\theta)]$ for the adsorbed molecules is the same for the two electrodes,
but even at $V_0=0$ there is an excess of outward (negative $cos(\theta)$) over
inward pointing OH bonds, which give rise to the potential drop between the
electrode and solution noted above. At different values of applied potential the
distribution of OH vectors is changed significantly.  At the positive electrode
(left panel of Fig. \ref{fig:OHorient}) the main effect of the electrode potential
is to deplete the population of OH bonds pointing into the electrode ($\cos(\theta)
< 0$).  At the negative electrode (right panel of Fig. \ref{fig:OHorient}) however,
at increased electrode potential there emerges a large population of OH vectors
which point into the electrode. This change in orientational structure at the
negative electrode is yet another demonstration of the asymmetry between the
solvent structure at the positive and negative electrode.

\begin{figure}
   \includegraphics[width=7cm]{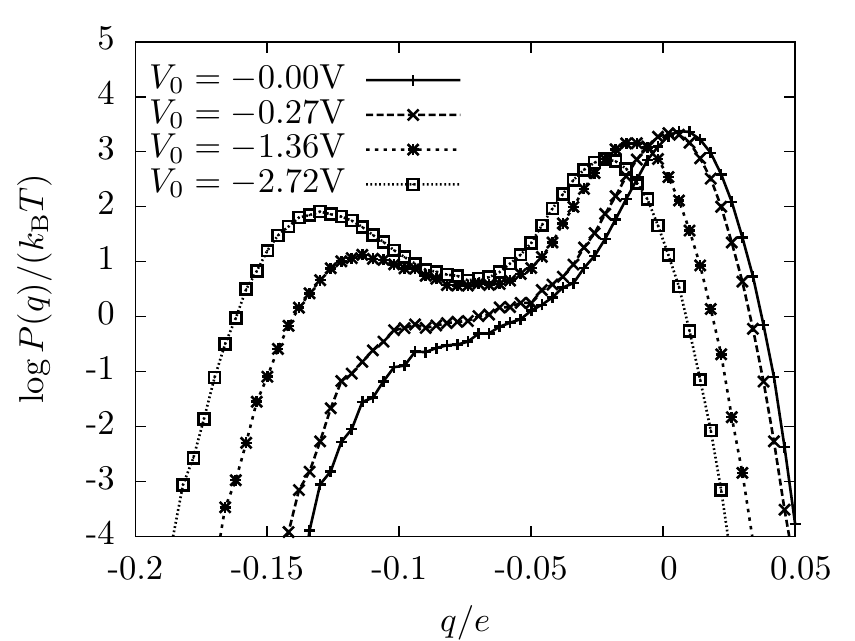}\includegraphics[width=7cm]{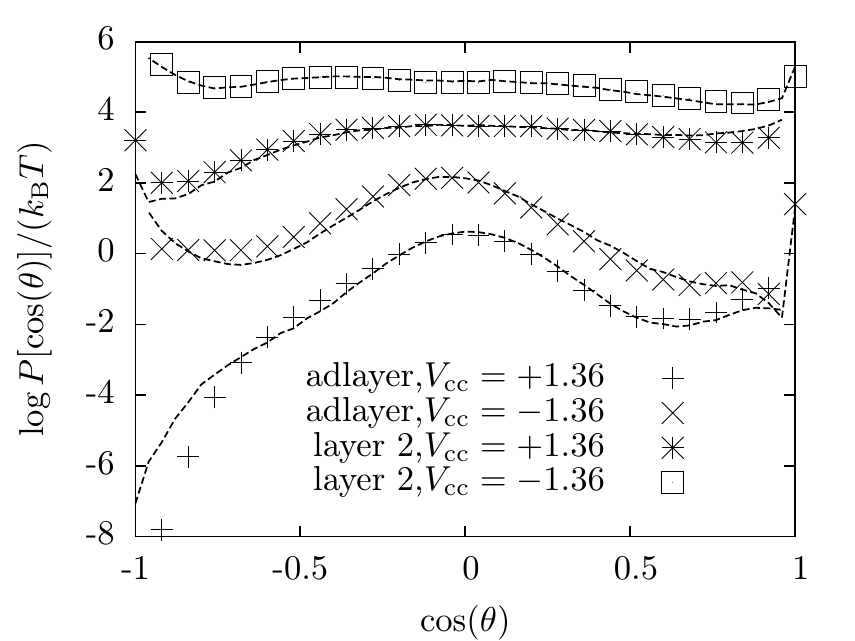}
\caption{The left-hand panel shows the distribution of the charges on the atoms
which make up the innermost layer of the negative electrode. The changes in the
orientations of the water molecules are accompanied by the development of a bimodal
distribution of wall charges. The right-hand panel contrasts the O-H bond
$\cos(\theta)$ distributions obtained in constant potential (figure
\ref{fig:OHorient}) and constant charge simulations; the former are shown by lines
and the latter by symbols. }
  \label{fig:wallcharge}
\end{figure}

The orientation of the adsorbed water molecules influences the charge induced on
the electrode atoms on which they sit. Figure \ref{fig:wallcharge} shows the
distributions of the charges induced on the atoms which make up the outermost layer
of atoms on the negatively charged electrode. At zero applied potential the average
charge is small and positive, for the reasons discussed above, but there is a
significant number of negatively charged atoms which are associated with an
adsorbed water molecule with an inward-pointing O-H bond. As the electrode
potential becomes increasingly negative, so does the mean charge, but the bimodal
character of the distribution becomes even more pronounced as more water molecules
flip an O-H bond towards the electrode.

One question which arises from these results is the extent to which they are
influenced by the inclusion of image charge interactions in the potential model. A
way of addressing this question is to carry out constant charge simulations in
which the values of the charges on the electrode atoms are fixed at the values of
the average charges on the first, second and third layers of the electrode atoms
obtained in constant potential runs with an electrode potential $V_0$. These static
charge distributions generate similar electrode potentials to the $V_0$ values used
in the constant potential runs used to generate them. It can be seen, from the
right-hand panel of figure \ref{fig:wallcharge}, that  the dynamical nature of the
charges in the constant potential simulations has only a small effect on the mean
orientational distributions of the water molecules in the adsorbed layer and none
on the second layer. However, as illustrated in figure \ref{fig:wallcharge}
left-hand panel, there is a local response of the electrode in the constant
potential simulations and this is responsible for the emergence of a significant
population of electrode-pointing OH bonds in the directly adsorbed molecules; it is
not observed in the simulations run at constant charge. The relatively small effect
of the image charges on {\it average} interfacial structure parallels the findings
in the molten salt simulations \cite{reed1}.
\begin{figure}
   \includegraphics[width=4.5in]{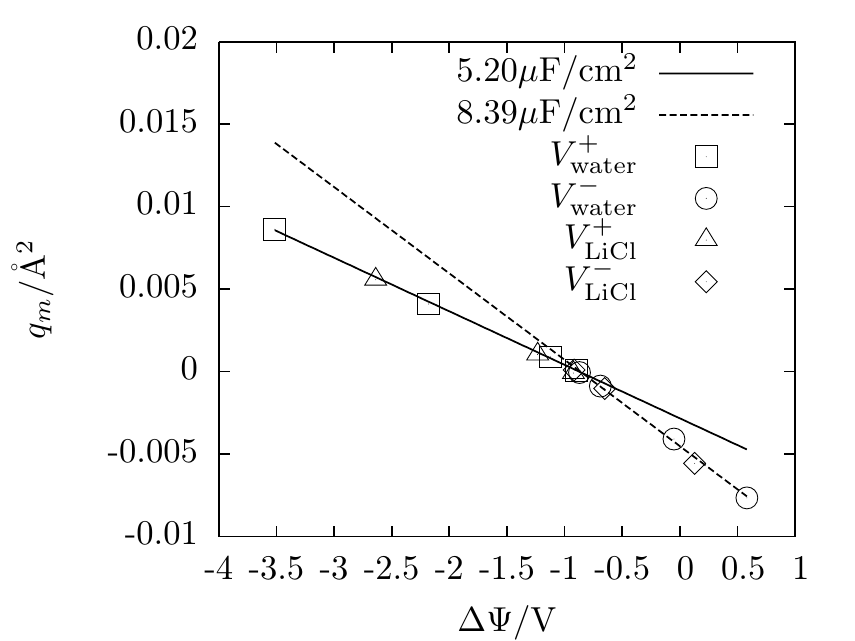}
\caption{The variation of the potential drop across the double layer, $\Delta \Psi$, with electrode charge density, $q_m$, for water and LiCl(aq) for the positive and negative electrodes.  The lines, whose slopes represent the differential capacitance of the positive and negative double layers are least squares fits to the water data.}
  \label{fig:capacitance}
\end{figure}

The layering of solvent and the molecular orientations within the layers adjacent
to the electrode affect the capacitance of the electrode, which can be measured
experimentally. The capacitance of the first two layers of solvent can be
calculated through the differential capacitance, $C = (\partial q_m/\partial \Delta
\Psi)$, where $q_m$ is the charge density on the electrode and $\Delta \Psi$ is the
potential drop across the first two layers of water. Figure \ref{fig:capacitance}
shows the dependence of $\Delta \Psi$ on $q_m$ for several values of the applied
potential. The plot reveals that the potential of zero charge (pzc) for our
simulated system is at -0.8 V. Comparing this quantity with experiment is not
straightforward, since in the experimental measurements the potential is quoted
with respect to a reference electrode whereas we can access directly the potential
difference between the interior of the electrode and the solution. The experimental
value with respect to a standard hydrogen electrode is +0.41 V \cite{bockris}. The
capacitance for electrode potentials which are on the positive side of the
potential of zero charge ($C=8.39 \mu \mathrm{F/cm^2}$) is larger than for negative
potentials, $C= 5.20 \mu \mathrm{F/cm^2}$. Both values are considerably lower than
predicted through experimental data which measures the the double layer capacitance
in the range of $C = 20-50 \mu \mathrm{F/cm^2}$ \cite{kolb,parsons} close to the
pzc. The substantial difference between our calculated value and experiment is not
surprising as our model does not include a realistic description of the electron
density at the metallic surface; the surface dipole potential arises from the
extension of the metal electrons into the interface beyond the nuclei and makes a
large contribution to the capacitance \cite{parsons}. This effect can be included
in a jellium model for the metal, as included in the theory of Schmickler and
Henderson \cite{schmickler}.

\section{Results for electrolyte solutions}
\label{ions}

\begin{figure}[htbp]
  \centering
 \includegraphics[width=4in]{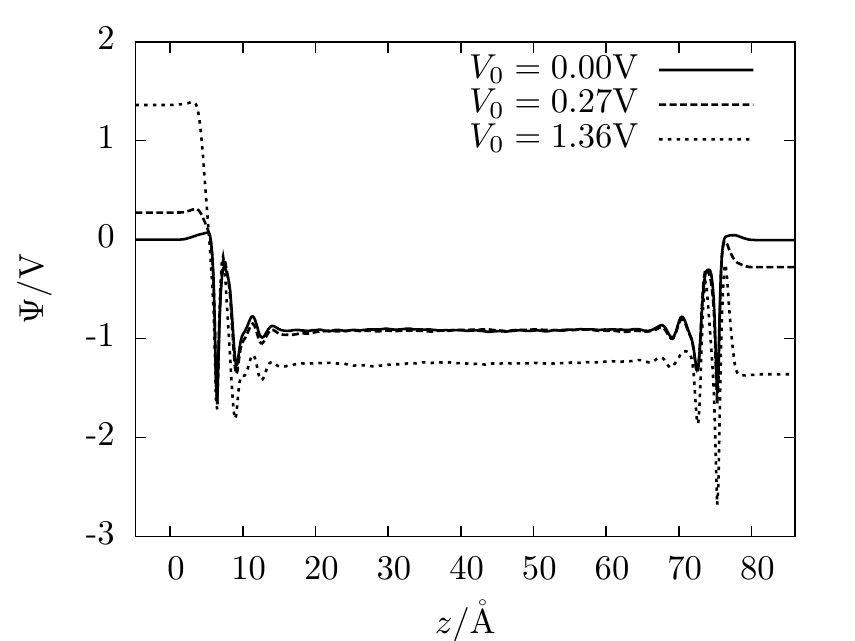}
\caption{The Poisson potential for a solution of LiCl(aq).  In contrast to the
Poisson profile for pure water (shown in Fig. 4), the LiCl solution screens the
electrodes perfectly beyond roughly 15 $\mathrm{\AA}$ from the electrode surface.}
\label{fig:poisson2}
\end{figure}

In figure \ref{fig:poisson2} we show the Poisson potential for an approximately one
molar solution of LiCl. In contrast to the pure water case (figure
\ref{fig:poisson1}) in the bulk region away from the interfaces, the Poisson
potential is now constant as a consequence of the screening by the ions present in
the solution. The oscillations in the potential across the interfacial region
closely resemble those in pure water at the same values of the applied potential.

\begin{figure}
\includegraphics[width=7cm]{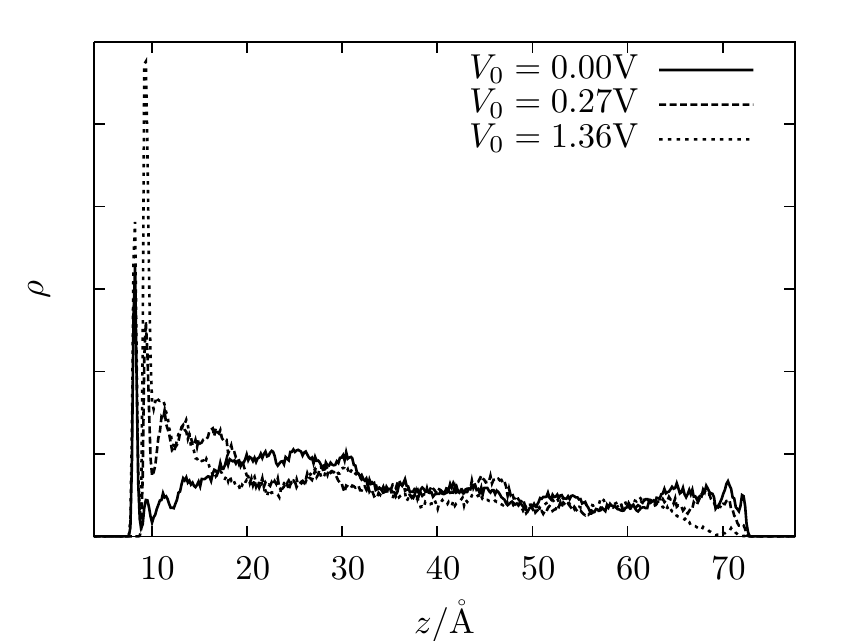}\includegraphics[width=7cm]{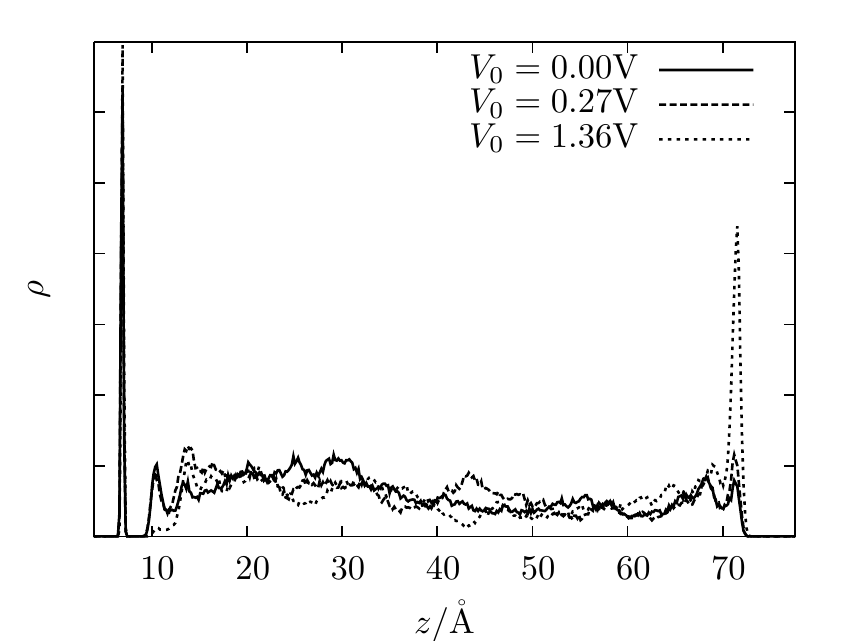}
\caption{The density profiles of the Li$^+$ and Cl$^-$ ions across the simulation
cells for various values of the applied potential. The results are the average of
runs of 2.5 nanoseconds, despite the lengths of these runs we see that
the average density profiles do not accord with those expected at full equilibrium.
Nevertheless, the Poisson potential profiles calculated in these runs do show
perfect screening (see figure \ref{fig:poisson2}).}
  \label{fig:iondensity}
\end{figure}

That the Poisson potential is exhibiting perfect screening is quite surprising as
the profiles of the ion density obtained by averaging over the whole simulation
runs are manifestly not well equilibrated (see figure \ref{fig:iondensity}). Even
at $V_0=0$ we see an excess of ions at the left-hand side of the cell, whereas the
equilibrium ion density profile should be constant, except close to the electrodes.
It would appear that efficient screening can be caused by an appropriate
\emph{local} arrangement of cations and anions, relaxation of the whole ion density
profile is not necessary.
The failure to reach a fully equilibrated ion density profile arises primarily
because of the slow rate of relaxation of the concentration by diffusion (and,
perhaps, an inappropriate initialisation of the ion positions in the simulations).
The rate should depend on the diffusion coefficient divided by the square of the
distance between the electrodes, and because we have used a large cell in the hope
of seeing the interfaces well separated by bulk, the relaxation times have become
extremely long. There is a second slow relaxation process, however. Examination of
the Li$^+$ profile close to the left-hand (positively charged) electrode shows a
sharp peak in the region associated with the adsorbed layer of water. This peak
arises from the presence of a single cation ion in this layer throughout the
$V_0=0$ and $V_0=0.27$ V runs; it was placed there in the initial configuration and
remained until the electrode potential was increased to 1.36 V. That this process
is so slow is because exchange of an ion between the strongly adsorbed layer of
water and the bulk is very slow, effectively because the ion cannot carry its
coordinating water molecules between the two regions.

We can examine the barrier which arises to prevent exchange between the adsorbed
layer and the bulk by umbrella sampling techniques. We compute the mean force,
$F(z)$, in the $z$ direction perpendicular to the electrode on an atom $i$ in the
simulation by constraining the atom $i$ at some position $z_0$ in a harmonic
potential $U_{z_0}(z) = {\frac {k}{2}}(z-z_0)^2$, where $k$ is the force constant
of the harmonic well, taken to be $100 ~\kB T \mathrm{\AA}^{-2}$. The mean force on
atom $i$ at position $z_0$ can be estimated as $F(z_0) = -k(\bar z - z_0)$ where
$\bar z$ is the average value of $z$ for species $i$ constrained to $U_{z_0}(z)$
during a simulation. We performed the mean force calculations on a $\mathrm{Li^+}$
ion and an oxygen centre of a water molecule in a LiCl(aq) solvent at zero applied
potential ($V_0 = 0.00$). In addition we computed $F(z)$ for the oxygen centre in a
pure water with electrode potentials $V_0 = \pm 2.72$ V. This method for generating
the mean force (and subsequently the potential of mean force (PMF) by integration)
can be sensitive to the set of initial conditions \cite{blue-moon}.  One set of
initial conditions, which corresponds to electrode desorption, was initiated by
choosing an already adsorbed species setting $z_0$ at the adsorption distance and
equilibrating with $U_{z_0}(z)$ for 1 picosecond.  The next member of this set of
initial conditions was created by setting $z_0 \rightarrow z_0+0.26 \mathrm{\AA}$
and again equilibrating with $U_{z_0}(z)$ for 1 picoseconds.  This process is
continued, in increments of $0.26 ~\mathrm{\AA}$ for approximately $5~
\mathrm{\AA}$.  Another set of initial conditions, corresponding to electrode
adsorption were generated in an analogous fashion by selecting an atom in the bulk
and moving $z_0$ towards the electrode in $0.26 ~\mathrm{\AA}$ steps.  The mean
force was computed by averaging $\bar{z}$ over a 20 picosecond trajectory.

\begin{figure}
   \includegraphics[width=7cm]{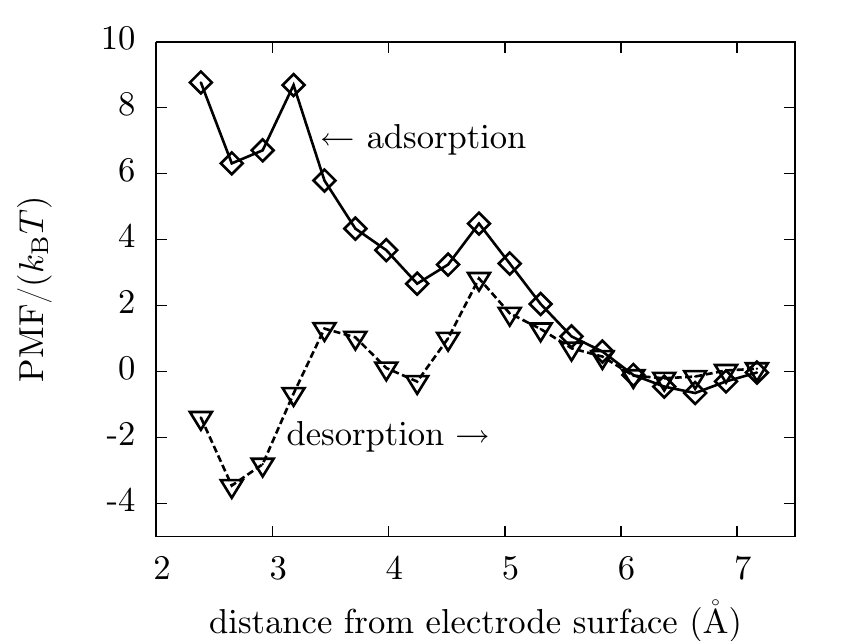}\includegraphics[width=7cm]{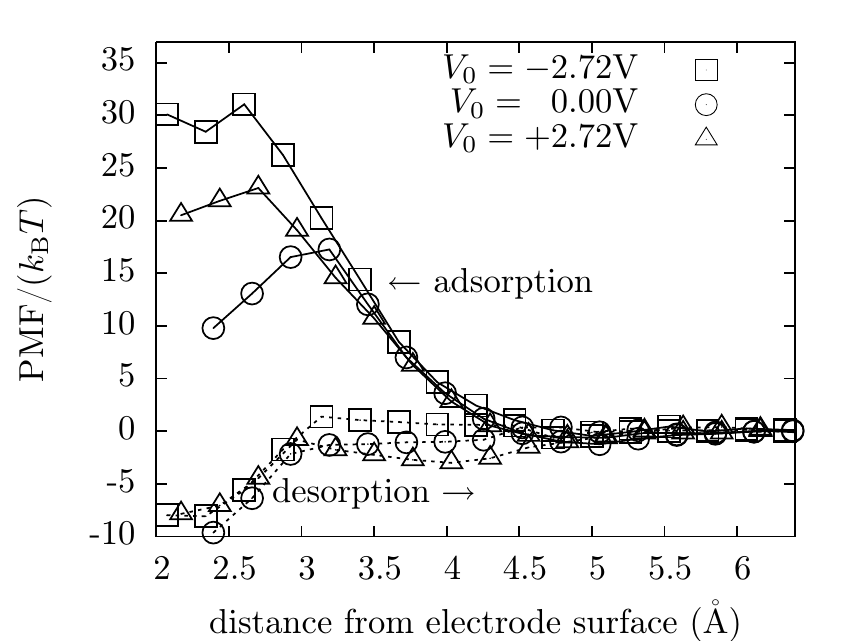}
\caption{The potentials of mean-force obtained for an Li$^+$ ion (left) and for the oxygen
atom (right) of a water molecule calculated with initial conditions appropriate to electrode desorption and electrode adsorption.  The curves with dash lines represent electrode desorption and the curves with solid lines represent electrode adsorption.  The results show a large degree of hysteresis, consistent with
fact that as a species is brought into the adlayer, a resident water molecule must
be displaced. }
  \label{fig:pmf}
\end{figure}

The potentials of mean force obtained by integration over $F(z)$ show a large
degree of hysteresis, which often arises when there is a free energy barrier  in
the chosen coordinate ($z$) frustrating equilibration on short timescales.  In
other words, the reaction mechanism for electrode adsorption is not correctly
characterized simply by a species distance from the electrode.  For the atom to
move into the adlayer it is necessary for an already adsorbed water molecule to
vacate an adsorption site, thus we might expect that a more suitable reaction
coordinate would describe the collective rearrangement which this entails.
Nonetheless, the calculated potential of mean force curves are informative.

If we focus firstly on the adsorption PMF for Li$^+$, we see that there is a
substantial barrier at about 5 \AA~ for the movement of the ion from the bulk into
a relatively stable position where the cation sits between the first and second
adlayers located about 4.4 \AA~ from the electrode surface; this position is
illustrated at the left-hand electrode in figure \ref{fig:cell}. This barrier
arises from the reorganisation of the solvation shell of the Li$^+$ ion which is
necessary for it to be accommodated in this layer. Note that a similar barrier
appears in the desorption pathway.  There is then a second barrier before the ion
is adsorbed at the electrode surface. In this region the hysteresis in the two
curves is pronounced. On the adsorption pathway the ion must force an already
adsorbed water molecule out of the way, so the energy increases steeply. On
desorption from the adlayer there is also a large energy increase as the process
leaves an empty adsorption site on the electrode surface.

The PMF for water shows no barrier for exchange of water molecules between the bulk
and the second adlayer. A large degree of hysteresis then sets in, associated with
the replacement of an already adsorbed water molecule by a molecule from the bulk.
The barrier to desorption from the first adlayer suggested by these data is of the
order of 10 $\kB T$, sufficient to lead to very slow exchange of the adsorbed water
and the bulk.

\section{Calculation of the Marcus curves for electron transfer.}

In order to examine how the electrical potential and the water structure in the
interfacial region affect the rate constants for electrochemical charge transfer we
have followed the scheme illustrated in figure \ref{fig:cell} for the aqueous
Ru$^{2+}$/Ru$^{3+}$ couple close to the model metallic electrode. Similar
calculations have been reported recently for a redox-active molten salt system
\cite{reed2}, where the problems caused by the very slow equilibration of the
concentration profiles we have noted above are not so marked and where the
statistical precision necessary to validate the calculations was relatively easily
obtained. We refer the reader to that paper for full details of the calculation and
merely recapitulate some essential details here.

We calculate the probability distribution functions for the vertical transition
energy between the two redox states, $\delta E_{{\rm Ru}^{2+} \to {\rm Ru}^{3+}}$
for oxidation and $\delta E_{{\rm Ru}^{3+} \to {\rm Ru}^{2+}}$ for reduction. The
vertical transition consists of changing the identity ({\it i.e.} charge and
interaction potentials) of a single ion at some configuration along an MD
trajectory with the electrode potentials set at some value $V_0$ and, without
changing the atomic positions (as befits the vertical or diabatic nature of the
Marcus curves), relaxing the electrode charges. The vertical transition energy is
the difference in the total interaction energy between the initial and final
states.  There is a constant term in this energy gap that depends upon the metal of
which the electrode is made (through its work function) and reacting ion (through
the gas-phase ionisation energy (Ru$^{2+}~\to$ Ru$^{3+}$+e$^-$), but independent of
the electrolyte solution. We have arbitrarily set the value of this constant to
make the mean energy gap of $\mathrm{Ru^{2+}}$ to be the negative of the that for
$\mathrm{Ru^{3+}}$ when $V_0 = 0$; the consequences of this will be illustrated
below.

\begin{figure}
\includegraphics[width=7cm]{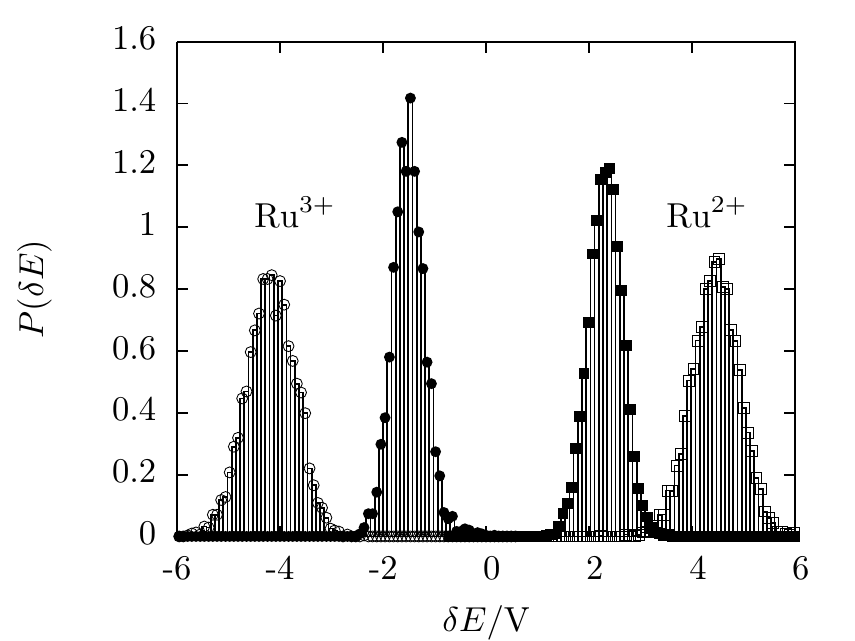}
\caption{The probability distributions $P_{{\rm Ru}^{2+}} (\delta E_{{\rm Ru}^{2+}
\to {\rm Ru}^{3+}})$ (squares) and $P_{{\rm Ru}^{3+}}(-\delta E_{{\rm Ru}^{3+} \to
{\rm Ru}^{2+}})$ (circles) at different positions in the cell.  The distributions
plotted with open symbols are taken in the bulk, i.e. $37~ \mathrm{\AA} \leq z \leq
39~ \mathrm{\AA}$, and the distributions plotted with filled symbols are taken
adjacent to the electrode, i.e. $71~ \mathrm{\AA} \leq z \leq 75~ \mathrm{\AA}$.}
\label{fig:probdist}
\end{figure}

In both oxidation and reduction, as discussed in detail in reference
\cite{reed2} a balancing charge is transferred to the electrodes and the energy of
this, which depends on the potential applied to the electrodes, is included in the
vertical transition energy \cite{footnote}. We then return the identity of the ion
to its initial value and continue the MD trajectory. By repeatedly sampling these
transition processes for redox species found at a given distance from the
electrodes we may build up probability distributions for the energy gaps, $P_{{\rm
Ru}^{2+}}(\delta E_{{\rm Ru}^{2+} \to {\rm Ru}^{3+}})$ and $P_{{\rm
Ru}^{3+}}(\delta E_{{\rm Ru}^{3+} \to {\rm Ru}^{2+}})$, at different positions in
the cell. Examples of the probability distributions are shown in figure
\ref{fig:probdist}. They are calculated for a sample of ions located in the middle
of the simulation cell, and compared to those of ions adjacent to the electrode.

The mean positions of the distributions and their widths are found to depend quite
strongly on the position of the redox ion in the cell, as we will discuss below.
The distributions are found to be rather accurately gaussian, which is the
expectation from Marcus theory if the surrounding medium responds linearly to the
change in the identity of the redox species. Our potentials describing the
interaction of the Ru$^{2+}$ and Ru$^{3+}$ with water were chosen so that both
cations had similar coordination shells and, as previous studies of redox processes
in the bulk have shown \cite{benjamin,sprik2,sprik_ru}, under these conditions it
is likely that the linear response limit is applicable.  Our data seems to be
consistent with linear response even when we consider the redox process for ions
close to the electrode surface, despite the strength of the interactions and the
restricted nature of the water molecules in the first adsorbed layer.

Following Sprik and co-workers \cite{sprik1,sprik2}, and making use of the special
properties of the mean vertical gap for oxidation ($\Delta E =\delta E_{{\rm
Ru}^{2+} \to {\rm Ru}^{3+}}=-\delta E_{{\rm Ru}^{3+} \to {\rm Ru}^{2+}}$) as a
reaction coordinate \cite{marcus2,warshel}, we may evaluate the free energies of
the Ru$^{2+}$ and Ru$^{3+}$ ions along this reaction coordinate from the
probability distributions
\begin{equation}
A_{{\rm Ru}^{2+}}(\Delta E)=-k_BT \ln P_{{\rm Ru}^{2+}}(\Delta E=\delta E_{{\rm
Ru}^{2+} \to {\rm Ru}^{3+}}) + {\bar A_{{\rm Ru}^{2+}}} \label{marcus1}
\end{equation}
and
\begin{equation}
A_{{\rm Ru}^{3+}}(\Delta E)=-k_BT \ln P_{{\rm Ru}^{3+}}(\Delta E=-\delta E_{{\rm
Ru}^{3+} \to {\rm Ru}^{2+}}) + {\bar A_{{\rm Ru}^{3+}}} \label{marcus2},
\end{equation}
where ${\bar A_{{\rm Ru}^{2+}}}$ corresponds to the free-energy at the minimum of
the Ru$^{2+}$ curve. Furthermore, when the vertical energy gap is taken as the
reaction coordinate the two free energy curves are linearly dependent, i.e.~\cite{tachiya,sprik2}
\begin{equation}
A_{{\rm Ru}^{3+}}(\Delta E)-A_{{\rm Ru}^{2+}}(\Delta E)=\Delta E. \label{marcus3}
\end{equation}
This apparently simple relationship is remarkably powerful; it means that we can
establish a relationship between the origins of the two curves (${\bar A}_{{\rm
Ru}^{3+}}$ and ${\bar A}_{{\rm Ru}^{2+}}$)  and also sample the free energy
surfaces for values of the reaction coordinate which are well away from the most
stable configurations simply by calculating the energy gap in the free-running
simulation. The ability to sample the curves away from their minima means that we
can obtain information on the Marcus curves in the vicinity of their crossing
point, which is the region which determines the kinetics of the electron transfer
event.

\begin{figure}
\subfigure[bulk ions, $37 ~\mathrm{\AA} \leq z \leq 39~ \mathrm{\AA}$]{
\includegraphics[width=7cm]{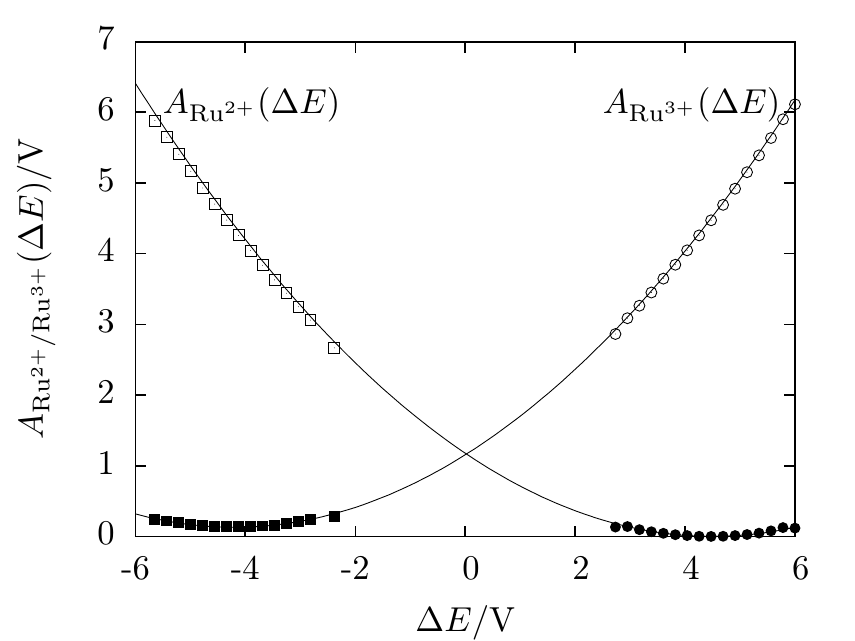}
} \subfigure[adsorbed ions, $72~ \mathrm{\AA} \leq z \leq 75 ~\mathrm{\AA}$]{
\includegraphics[width=7cm]{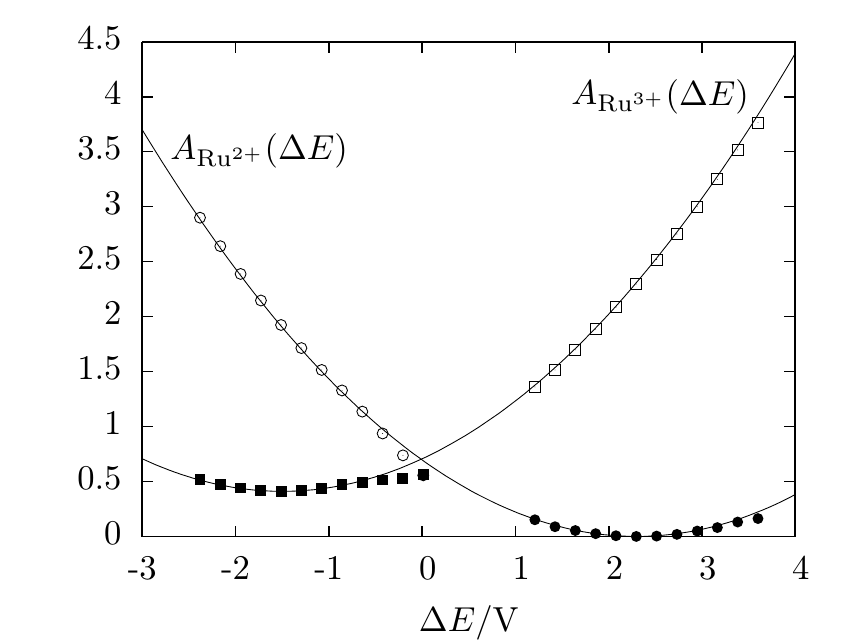}
} \caption{The free energies of $\mathrm{Ru^{2+}}$ (squares),  and
$\mathrm{Ru^{3+}}$ (circles) along the reaction coordinate $\Delta E$.  The filled
symbols are the distributions as computed from simulation data and the hollow
symbols are constructed using the relationship in Eq. \ref{marcus3}.  Solid lines
are the linear response (Marcus) prediction, computed using $\langle \Delta E_1
\rangle$ and $\langle \Delta E_2 \rangle$ along with Eqs. 7-10.  Panel (a) is a
representative distribution representative of bulk ions and panel (b) is the
distribution of ions adsorbed at the electrode. } \label{fig:marcuscurves}
\end{figure}

The data points obtained from (\ref{marcus1})- (\ref{marcus3}) are plotted in
figure \ref{fig:marcuscurves} for  $V_0=0$. Note that our choice of the arbitrary
energy added to the gap to represent the work function and ionisation energy has
resulted in only a small difference between the mean free energies of the oxidised
and reduced forms for the mid-cell position: experimentally, the reduction
potential for this couple is 0.249 V with respect to the standard hydrogen
electrode, so the relative positions of the minima in the curves should be similar
to reality and the electron transfer in the ``normal" Marcus r\'egime at $V_0=0$.

If the probability distributions really are Gaussian, equations \ref{marcus1} and
\ref{marcus2} show that the Marcus free-energy curves will be harmonic about the
mean values of the reaction coordinate for the oxidation and reduction processes,
{\it i.e.} the peak positions of the respective probability distributions $\langle
\Delta E_1\rangle$ and $\langle \Delta E_2 \rangle$, respectively. It was shown by
Tachiya \cite{tachiya} that under this Gaussian assumption all properties of the
Marcus curves can be {\em predicted} simply from a knowledge of $\langle \Delta
E_1\rangle$ and $\langle \Delta E_2 \rangle$; the necessary manipulations are
described in the previous paper \cite{reed2,sprik2}. These predicted curves are
shown by solid lines in figure \ref{fig:marcuscurves} and are seen to provide an
accurate representation of the data. This applies both for the data obtained for
redox ions close to the centre of the cell and also close to the electrode
surfaces, despite the fact that the values of $\langle \Delta E_1\rangle$ and
$\langle \Delta E_2 \rangle$ themselves depend quite strongly upon the distance
from the electrode. The position-dependence of the widths of the probability
distributions which we noted in discussing figure \ref{fig:probdist} is therefore
seen to be contained within the Gaussian description of the fluctuations in the
reaction coordinate and related to the position-dependence of $\langle \Delta
E_1\rangle$ and $\langle \Delta E_2 \rangle$.

\begin{figure}
\includegraphics[width=7cm]{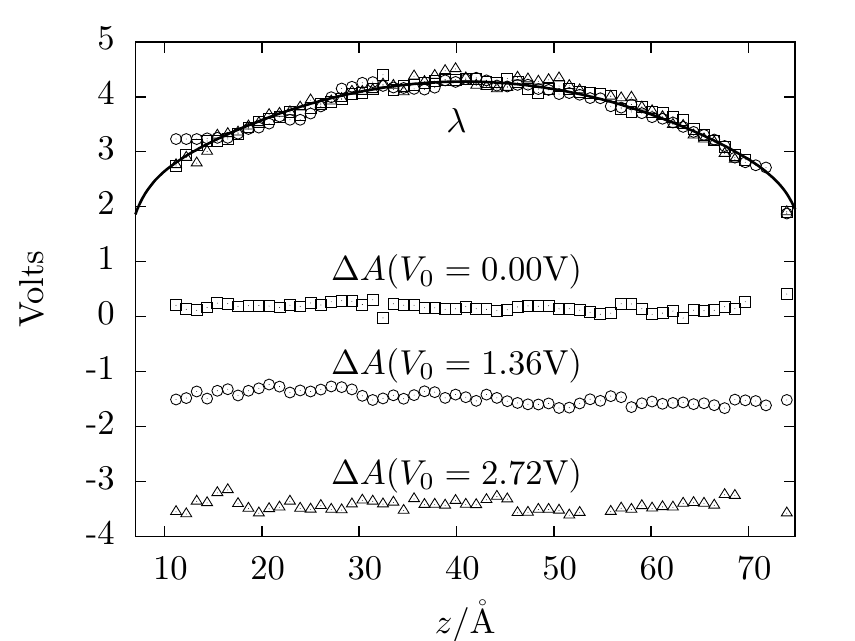}
\caption{Dependence of the reaction free energy $\Delta A$ (lower three curves) and
the reorganization energy $\lambda$ (upper three curves) on cell position at
different values of electrode potential $V_0$.  The solid lines shows the variation
of the energy (shifted by an additive constant) of a test charge as it is moved
across an otherwise empty cell, which reflects the energy associated with the image
charge effects in our simulations.} \label{fig:DeltaAandlambda}
\end{figure}

The parameters which are normally used to describe  the shapes of the Marcus curves
are $\Delta A $ and the reorganization energy $\lambda$, see figure
\ref{fig:marcustheory}. In the Gaussian/linear response r\'egime, both may be
written in terms of the mean energy gaps \cite{tachiya,sprik2}:
\begin{equation}
\label{eqn:deltaA}
  \Delta A = {\frac {1}{2}}(\langle
\Delta E_1 \rangle + \langle \Delta E_2 \rangle)
\end{equation}
and
\begin{equation}
\label{eqn:lambda}
  \lambda = {\frac {1}{2}}(\langle
\Delta E_1 \rangle - \langle \Delta E_2 \rangle),
\end{equation}
with $\lambda=\lambda^\prime$ in figure \ref{fig:marcustheory}. In this re\'gime
the activation free energy for electron transfer is given by the famous expression
\cite{marcus1}
\begin{equation}
\Delta A^\ddag={\frac {(\Delta A + \lambda)^2}{4\lambda}}.
\end{equation}
The dependence of $\Delta A$ and $\lambda$ on the position of the redox ion in the
cell and on the applied potential is illustrated in figure
\ref{fig:DeltaAandlambda}. The behaviour of these parameters parallels that seen in
the molten salt simulations \cite{reed2} and we refer to that paper to fully
vindicate the interpretations of the data which we offer below.

The reorganisation energy $\lambda$ is seen to be virtually independent of the
applied potential, but strongly dependent on the position of the ion in the cell.
The latter is associated with the way in which the polarization of the electrodes
(image charge effect) contributes to the vertical energy gap. In passing from the
initial state, say Ru$^{2+}$, to the final state Ru$^{3+}$ we create a unit
positive charge at the location of the redox ion. When we allow the relaxation of
the electrode charges to re-establish the constant potential condition we not only
allow the transfer of one unit of negative charge to the electrodes, we also allow
the electrode to be polarized by the newly-created positive charge. The interaction
between the newly created image charge and the change in the charge of the redox
ion {\em is not screened} because the positions of the electrolyte atoms do not
relax after the excitation event in a diabatic description of the charge transfer
process. Since the image effect contributes to $\langle \Delta E_1 \rangle$ and $
\langle \Delta E_2 \rangle $ with equal magnitude but opposite sign, it does not
affect the value of $\Delta A$, which is seen to be $z$-independent. The
reorganization energy is, however, strongly affected by the image effect. Marcus
\cite{Marcus_image} obtained an expression for the reorganization energy
appropriate to an ion in a dielectric fluid at a distance $d$ from a single
metallic surface,
\begin{equation}
\lambda_{Marcus}(d) = \delta q^2\left[ {\frac {1}{\varepsilon_{\infty}}}-{\frac
{1}{\varepsilon_s}} \right] \left({\frac {1}{2a}}-{\frac {1}{2d}}\right),
\label{reorganisation}
\end{equation}
where $\delta q$ is the charge difference between the reduced and oxidised species.
When $d$ is large, this expression gives the reorganization energy for a redox
process in the bulk fluid: it contains the contribution the non-electronic part of
the dielectric response of the fluid ({\it i.e.} that caused by reorganization of
the nuclear positions) to the change in the charge of a redox species with radius
$a$; as such it involves (in the first bracket) the difference between the static
$\varepsilon_s^{-1}$ and infinite frequency $\varepsilon_{\infty}^{-1}$
longitudinal dielectric susceptibilities. We cannot compare directly with this
expression because our sample geometry has {\em two} metallic surfaces and is
periodic in the transverse direction. However, we can compare directly with the
position-dependent energy of a charge introduced into an empty simulation cell,
this is the effective image interaction energy in our periodic system \cite{reed2}
when the newly created charge is in a vacuum. Away from the interfaces, any
difference between this quantity and the reorganisation energy should reflect the
effective dielectric screening function of our simulated electrolyte ({\it i.e.}
the factor analogous to the square-bracketed term in equation
\ref{reorganisation}). In fact, we see that the two curves coincide well showing
that the factor is indistinguishable from one. In our simulated system the water
molecules and ions are not polarizable, so $\varepsilon_{\infty}$ is just unity and
since for SPC/E water $\varepsilon_s$ is about 70 we can only conclude that our
data is consistent with the Marcus expression. Close to the electrodes, the
reorganisation energy does appear to depart from the modified Marcus expression,
and this could be associated with the effect of the proximity of the electrode on
the solvation characteristics of the water molecules there. However, the statistics
in this domain are not good, as the ruthenium ions are even more reluctant to
reorganise their solvation shells and approach the electrode than were the Li$^+$
ones.  Better sampling methods for the vertical energy gaps are required before
firm conclusions may be drawn.

\begin{figure}
\includegraphics[width=7cm]{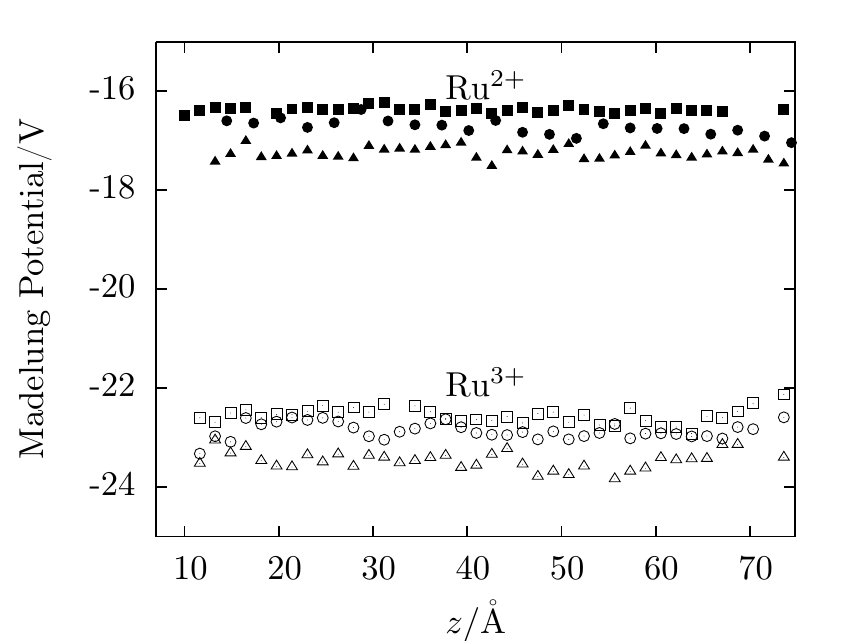}
\caption{Variation of the Madelung potential across the simulation cell for the $\mathrm{Ru^{2+}}$ (filled symbols) and $\mathrm{Ru^{3+}}$ (hollow symbols) ions at different values of the electrode potential, $V_0 = 0.00V$ (squares), $V_0 = 1.36V$ (circles), and $V_0 = 2.72V$ (triangles).  The Madelung potential, which is the potential felt by a species, does not exhibit the kinds of oscillations seen in the Poisson potential (Fig. \ref{fig:poisson2}) in the vicinity of the electrode.}
\label{fig:madpotprof}
\end{figure}

In the central part of the simulation cell, the reaction free-energy $\Delta A$
varies linearly with the potential applied to the electrode to which the electron
is transferred. This reflects the change in the energy of the electron which is
transferred to the electrode, which, as we have emphasised, contributes to the
free-energy of the oxidised state. As the redox species approaches the electrode
surface, the reaction free-energy seems to be remarkably constant. It might have
been expected to show the kind of fluctuating behaviour evident in the Poisson
potential, since conventional electrostatic considerations would suggest that this
potential should influence the relative energies of the doubly and triply charged
ions. However, as discussed in the molten salt context \cite{reed2},  this is
\emph{not} the potential which should be used to discuss the changes in the energy
levels of an ion. Rather, we should be considering the potential at the ion's
centre due only to {\em the other} charges present in the system: this might be
better called a Madelung potential. The difference between the two potentials is
surprisingly large, as illustrated in figure \ref{fig:madpotprof} where we show the
$z$-dependence of the mean Madelung potentials experienced by the Ru$^{2+}$ and
Ru$^{3+}$ ions compared with the Poisson potential.

Not only does the Madelung potential depend on the identity of the  species on
which the potential is evaluated, it is seen to be constant across the simulation
cell except in the immediate vicinity of the electrode surfaces - its behaviour
illustrates perfect screening much closer to the electrode surface. The
$z$-independence of the Madelung potential therefore provides a much better
explanation of the insensitivity of $\Delta A$  to $z$ than does the Poisson
potential.
\section{Summary and Conclusion}
The methods described have allowed a full, self-consistent calculation of the
liquid structure and electrical potentials for an aqueous ionic solution close to a
model metallic wall maintained at a constant electrical potential. The simulation
is a direct realisation of a model electrochemical interface, as it appears in text
books. Using a realistic potential for water-platinum interactions, we find a
strongly absorbed layer of water molecules on the electrode with the molecules
oriented in the plane of the interface at zero potential, in common with earlier
studies \cite{berkowitz}. Despite the strength of the absorption, the water
molecules do reorient as the electrode potential is changed and this affects the
behaviour of the electrical potential across the interface and the differential
capacitance of the electrode. The absorption of cations at the electrode is
strongly inhibited by the requirement for them to reorganise their hydration shells
to approach the electrode surface.

We have begun to characterise how the interfacial water affects the rate constant
for electrochemical charge transfer by directly calculating the Marcus free energy
curves for the oxidised and reduced species at different positions in the cell with
a particular choice of reaction coordinate. The fluctuations in the solvation
structure which influence these curves were shown to be accurately Gaussian for the
modelled Ru$^{2+}$/Ru$^{3+}$ couple, consistent with linear response of the solvent
to the charge state of the redox ion. The reorganisation energy was strongly
dependent on the distance of the redox species from the electrode surface and
independent of the electrode potential. The effect was traced to image charge
interactions with the metal surface. With the statistics available to us at
present, we could not detect an effect of the altered dynamical characteristics of
the absorbed water on the solvation fluctuations when the redox species was close
to the electrode surface. The reaction free energy $\Delta A$ measures the
difference in the free energies of the oxidised and reduced states with the redox
ion at a given distance from the electrode surface. Contrary to textbook
expectations, its position dependence does not resemble the mean electrical
potential. It only deviates from the bulk value in the immediate vicinity of the
interface where the competition between solvating the electrode and solvating the
redox species becomes a factor.

\section{\label{sec:acknowledgments}Acknowledgments}
This work was supported by EPSRC, through Grant GR/T23268/01, as well as by the
Director,  Office of Science, Office of Basic Energy Sciences, Chemical Sciences,
Geosciences, and Biosciences Division, U.S. Department of Energy under Contract No.
DE-AC02-05CH11231.

\bibliographystyle{unsrt}

\begin{thebibliography}{1}
\bibitem{footnote} As discussed at length in reference \cite{reed2},
 because of the adiabatic nature of our calculations with respect to the electrode charges,
a fraction of the charge is transferred to each electrode, whereas in the normal
description of electrochemical charge transfer we envisage a whole electron being
transferred to a designated electrode.  The value of the fraction depends
systematically on the position of the ion in the cell and we may therefore correct
our vertical transition energies to correspond to electron transfer to the anode,
as decsribed in detail \cite{reed2}. We therefore quote our vertical gap energies
as if the electron transfer event is occurring at the anode.
\end{thebibliography}

\end{document}